\documentclass[twocolumn,preprintnumbers,amsmath,amssymb]{revtex4}
\usepackage{graphicx}
\usepackage{dcolumn}
\usepackage{amsmath}
\usepackage{bm}
\DeclareGraphicsExtensions{.pdf,.png,.jpg}
\usepackage{epstopdf}
 
\begin{document}
 
\title{Electronic structures of doped BaFe$_2$As$_2$ materials: virtual crystal approximation versus super-cell approach}
\author{Smritijit Sen, Haranath Ghosh}

\affiliation{Indus Synchrotrons Utilization Division and Homi Bhabha National Institute, Raja Ramanna Centre for Advanced Technology, 
Indore -452013, India. \\}
\date{\today}
\begin{abstract}
Employing virtual crystal approximation and super-cell methods for doping, we have performed 
a comparative study of the electronic structures of various doped BaFe$_2$As$_2$ materials 
by first principles simulations. Both of these methods give rise to a similar density of states 
and band structures in case of hole doping (K doping in Ba site) and iso-electronic P doping 
in As site. But in case of electron doped systems with higher doping concentration, electronic structures, 
calculated using virtual crystal approximation approach deviates from that of the super-cell method.
On the other hand in case of iso-electronic Ru doping implemented by virtual crystal approximation, an 
extra shift of the chemical potential in electronic structure in comparison to super-cell method 
is observed and that shift can be used to predict the correct electronic structure within virtual 
crystal approximation as reflected in our 
calculated Fermi surfaces. But for higher Ru doping concentration, simple shifting of chemical potential 
does not work as the electronic structure calculated by virtual crystal approximation approach 
is entirely different from that of the calculated by super-cell formalism. 
\vspace{1pc}
\end{abstract}
\maketitle

\section{Introduction}
Advent of high temperature superconductivity in Fe-based compounds in the vicinity of magnetism
\cite{Kamihara,Cruz,Dong} has led to immense intrigue among the researchers to resolve the unsolved mysteries
in the field of high temperature unconventional superconductivity. Apart from some old issues like 
superconducting pairing symmetry, proximity of superconductivity and magnetism, unusual 
Fermi surface (FS) topology, a number of new phenomena like existence of orbital density wave, 
origin of nematic phase {\it etc.,} are still under substantial debate 
\cite{Mazin2011,Hirschfeld,Mazin2015,Lumsden,Marsik,Jalcom,Pengcheng,Shimojima,Kontani,Ghoshepl,Baek,Fernandes}. 
In literature there
are large number of experimental as well as theoretical studies regarding the electronic structures
and properties of Fe-based superconductors \cite{Kontani,Mazin,Chubukov1,Onari,Stewart} to establish the mechanism of 
unconventional superconductivity with high T$_c$. It should be mentioned that 
the highest transition temperature (T$_c$) till now for Fe-based superconductors is 109K \cite{Ge}.
Correlation of superconducting transition temperature and other physical properties with 
structural parameters is well established \cite{Mizuguchi,Lee,Zhao,Shirage,Acta,sust}. 
Fe-As(pnictides)or Fe-Se(chalcogenides) planes are back bone structures of these 
superconductors where As atoms are placed above and below the Fe plane.
Height of this As atom from Fe plane is defined as 'anion height' closely related to z$_{As}$ {\it i.e.,}
the fractional co-ordinate of As atom.
It is also evident from literature that, 
as far as veracity of calculation of electronic structure is concerned, the role of structural 
parameters specially z$_{As}$ is instrumental \cite{zAs,Acta,sust}.  
Fe-based superconductors are broadly classified into 
six families among them 122 family is the most studied family, both experimentally as well as theoretically.
The main reason for that is the availability of high quality single crystal for 122 systems.
It should also be noted that the parent compound BaFe$_2$As$_2$ is not superconducting in ambient 
pressure but it shows magnetic order in the form of spin density wave (SDW). Superconductivity appears in this system when 
 SDW order is suppressed and that can be done either by applying 
hydrostatic pressure \cite{Mani,Park,Alireza} or chemical pressure {\it i.e.,} by doping.
Superconductivity can arise into the system by doping on any of the three sites including Fe sites \cite{Sefat,Jasper,Liet}.
Another interesting case is isoelectronic Ru substitution in place of Fe which also forge the
system to become a superconductor \cite{Sharma1,Fan}. Electronic structure of all these doped 122 systems 
has been studied using first principles calculation based on density functional theory (DFT)
with a great success \cite{Wang,DJSingh,Zhang}. But there are some challenges regarding the application of DFT in 
Fe-based superconductors whose electronic structure is very much sensitive to temperature, 
pressure and doping. We extend this issue in the theory and computational method section in a greater detail. 

As mentioned earlier in BaFe$_2$As$_2$ system, one can dope on any of the sites to make it superconducting.
Phase diagrams of Fe-based superconductors play a crucial role in the understanding of various physical 
properties, which are very sensitive to doping and temperature. A generic phase diagram of 122 family of 
Fe-based superconductors consists of various phases like superconductivity, SDW order, C$_2$ and C$_4$ structural transition, 
nematic phase {\it etc.} \cite{Avci,Lucarelli,XFWang,Ni,Thaler,Nandi,Kasahara}. All 
these phases are intimately dependant on doping concentration. There exists a large diversity in the phase diagrams
of various doped BaFe$_2$As$_2$ systems. For electron doped system superconducting and other 
exotic phases exist in the system upto 15-20$\%$ doping concentration which is very less 
compared to other 122 systems \cite{Ni,Thaler,Nandi}.
 Theoretically this doping effect can be accomplished by several ways but two of them are widely used: super-cell (SC) method
 and virtual crystal approximation (VCA). We later discuss these two methods concisely.
 Application of these two methods to implement doping for calculating electronic structures 
 (density of states, band structures,
 Fermi surfaces {\it etc.}) of various doped 122 systems are there in the literature.
 But a comparative study of electronic structures calculated using both these two approaches is rare.
 In this work we proclaim the differences in electronic structures of the various doped 122 systems
 calculated by implementing doping in VCA and SC methods.  
 Five doped BaFe$_2$As$_2$ systems {\it viz.} Ba$_{1-x}$K$_x$Fe$_2$As$_2$ (hole doping), 
 BaFe$_{2-x}$Co$_x$As$_2$ (electron doping), BaFe$_{2-x}$Ni$_x$As$_2$ (electron doping), 
 BaFe$_2$(As$_{1-x}$P$_x$)$_2$ (isovalent doping) 
 and BaFe$_{2-x}$Ru$_x$As$_2$ (isovalent doping) has been examined through detail study of 
 density of states and band structures using VCA and SC approaches.
 In particular, we observe that both these methods give rise to very different electronic structures
 when one dope on Fe site. We also found that for higher doping concentration 
 electronic structures calculated using VCA method deviate from that of the calculated utilizing
 SC formalism for certain cases. Our work demonstrate a detailed study of electronic structures of all 
 these above mentioned doped 122 systems where doping has been implemented by two different theoretical formalisms.
 In order to assert a correct electronic structures of these doped systems one must assure the validity of 
 VCA formalism and one should also be aware of the limitation of these two methods.
  \begin{figure}[ht]
    \centering
   \includegraphics [height=5cm,width=6cm]{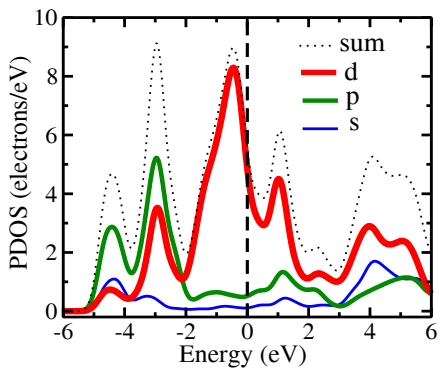}
    \caption{(Colour online) Calculated orbital projected partial density of states (PDOS) of 
    undoped BaFe$_2$As$_2$ system within the energy range
    -6 eV to 6 eV. PDOS of various orbitals like s, p, d and total density of states are indicated by 
    blue, green, red solid lines and black dotted line (sum) respectively. Vertical dotted line at 0 eV represents the
    Fermi level.}
   \label{PDOS_undoped}
  \end{figure}
  \begin{figure}[ht]
    \centering
    \includegraphics [height=7cm,width=8cm]{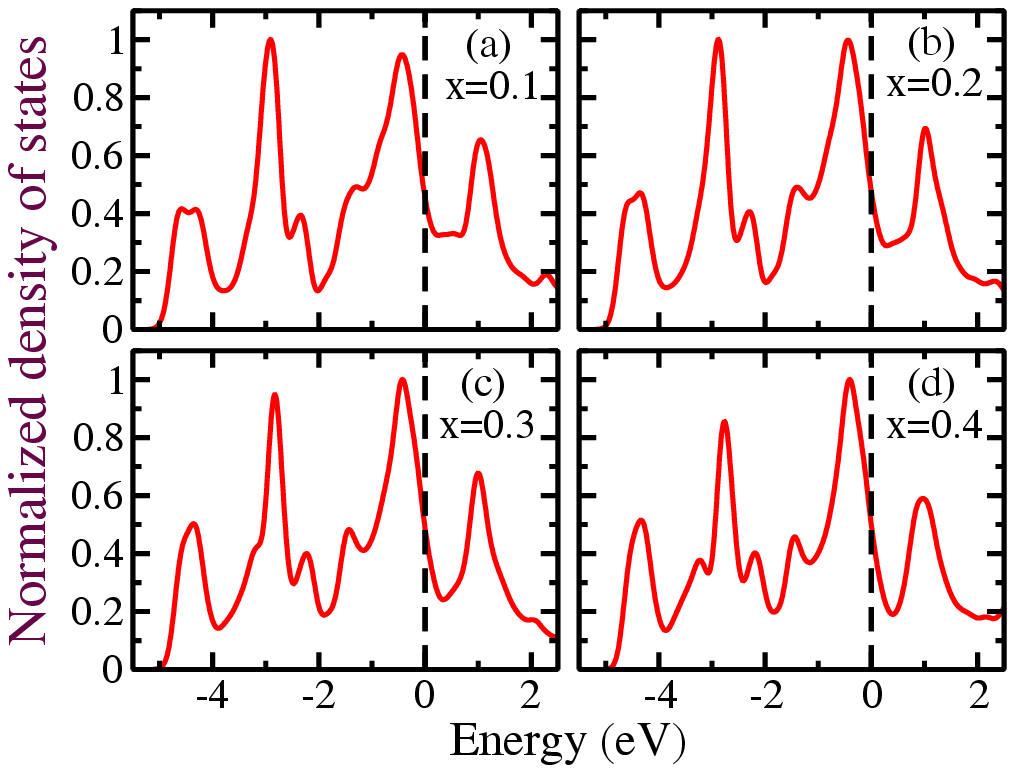}
    \caption{(Colour online) Calculated normalized total density of states (red solid line) using VCA approach
    for Ba$_{1-x}$K$_x$Fe$_2$As$_2$ (hole doping) system with (a) $x=0.1$, (b) $x=0.2$, (c) $x=0.3$ and (d) $x=0.4$
    within the energy interval -5.5 eV to 2.5 eV (near Fermi level, indicated by vertical dotted line at 0 eV).}
   \label{K_dos_vca}
  \end{figure}
  \begin{figure}[ht]
    \centering
    \includegraphics [height=7cm,width=8cm]{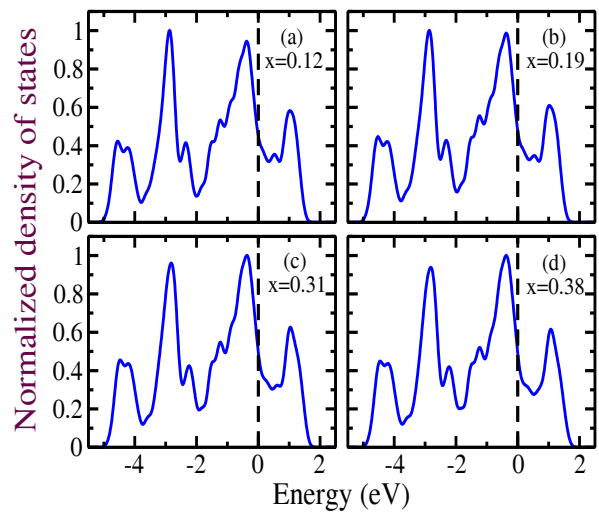}
    \caption{(Colour online)Calculated normalized total density of states (blue solid line) using SC approach
    for Ba$_{1-x}$K$_x$Fe$_2$As$_2$ (hole doping) system with (a) $x=0.12$, (b) $x=0.19$, (c) $x=0.31$ and (d) $x=0.38$
    within the energy interval -5.5 eV to 2.5 eV (near Fermi level, indicated by vertical dotted line at 0 eV).}
   \label{K_dos_sc}
   \end{figure}
\section{Theory and Computational Method}
Our first principles electronic structure calculations were accomplished by using 
plane wave psudopotential method based on density functional theory \cite{CASTEP},
where the electronic exchange correlation is treated under the
 generalized gradient approximation (GGA) using Perdew–
Burke–Ernzerhof (PBE) \cite{PBE}/PW91 \cite{PW91} functional. A number of authors addressed to the fact that density funtional 
theory within local density approximation (LDA) as well as generalized gradient approximation (GGA) was
 unable to achieve the experimental value of z$_{As}$ (fractional co-ordinate of As)
 fairly accurately \cite{Acta,Mazin,sust,DJSingh,Zhang,Yin}. In fact the optimized value of z$_{As}$ 
 is about 0.1 {\AA} smaller than that of the experimental z$_{As}$. The reason behind this discrepancy 
 is the presence of strong magnetic fluctuation, emerging due to the existence of Fe atoms
 in these materials \cite{Mazin2}. Also it should be noted that electronic structures (like Fermi surface),
 calculated using optimized lattice parameters like a, b, c and z$_{As}$ do not resemble with that of the 
 experimentally measured one \cite{pla,zAs}. All these studies force us to take experimental 
 lattice parameters {\it i.e.,} a, b, c and z$_{As}$ \cite{thesis,Acta,Sefat,Ni} instead of geometry optimized 
 (total energy minimization) lattice parameters as the inputs of our first principles 
 electronic structure simulations. We use room temperature tetragonal experimental 
 lattice parameters of undoped as well as various doped BaFe$_2$As$_2$ systems to simulate the
 electronic structures (band structure, density of states). 
 Non-spin-polarized single point energy calculations were carried out for tetragonal phase with
 space group symmetry I4/mmm (No. 139) using ultrasoft pseudo-potentials and plane-wave basis 
 set with energy cut off 500 eV and higher
 and self-consistent field (SCF) tolerance as $10^{-6}$ eV/atom. 
 Brillouin zone is sampled in the k space within Monkhorst–Pack scheme
 and grid size for SCF calculation is chosen as per requirement of the calculation 
 for various systems and approaches. For simulating Fermi surfaces, grid size of SCF calculation is chosen as
 $30\times 30\times 41$.
 As mentioned earlier in this work we theoretically implement the 
 doping using two different approaches (a) Virtual crystal approximation (VCA) and (b) Super-cell (SC) method.
 Below we briefly discuss the theory, development, advantages and disadvantages of the two formalisms.
 
 Study of disordered systems employing first principles electronic structure methods 
 requires some amount of approximation to implement disorder or doping theoretically.
 To treat such disordered or doped systems there exist two very popular methods as mentioned earlier:
 super-cell (SC) method and virtual crystal approximation (VCA). It is very important to mention the 
 advantages and shortcomings of both the approaches before we  
 discuss the theoretical background of those methods. The super-cell method can give 
 more accurate results than the other one but certainly needs much more computational resources 
 as well as time compared to the VCA formalism. Accuracy of calculation is associated to the
 fact that the SC method can portray the local interaction between two 'real' atoms 
 whereas the VCA method unable to recount that. 
 One of the most important drawback for SC method is the artificially 
 imposed periodicity which sometime may not mimic the actual 
 physical situation of randomly doped system. This approach certainly .
 demand for a critical look at the finite size effect as well as the imposed artificial periodicity.
 On the other hand, a much lucid and 
 computationally inexpensive approach for treating doped systems is to introduce the virtual
 crystal approximation \cite{Nordheim}, in which a doped crystal with the original periodicity, 
 but composed of fictitious ‘‘virtual’’ atoms is created to mimic the actual doped system.
 To apply VCA method in order to study the case of doping, one must concern about certain important 
 issues: firstly the accuracy of calculation, secondly the efficiency for treating the heterovalent systems
 and lastly the electronic property that we are trying to calculate. There exists a different 
 formalism called ‘‘computational alchemy’’ \cite{Marzari, Saitta} to go beyond VCA. But 
 this method is much more intricate than slandered VCA formalism, demanding
 the utilization of density-functional linear-response methods.
 \begin{figure}[ht]
  \centering
  \includegraphics [height=7cm,width=8cm]{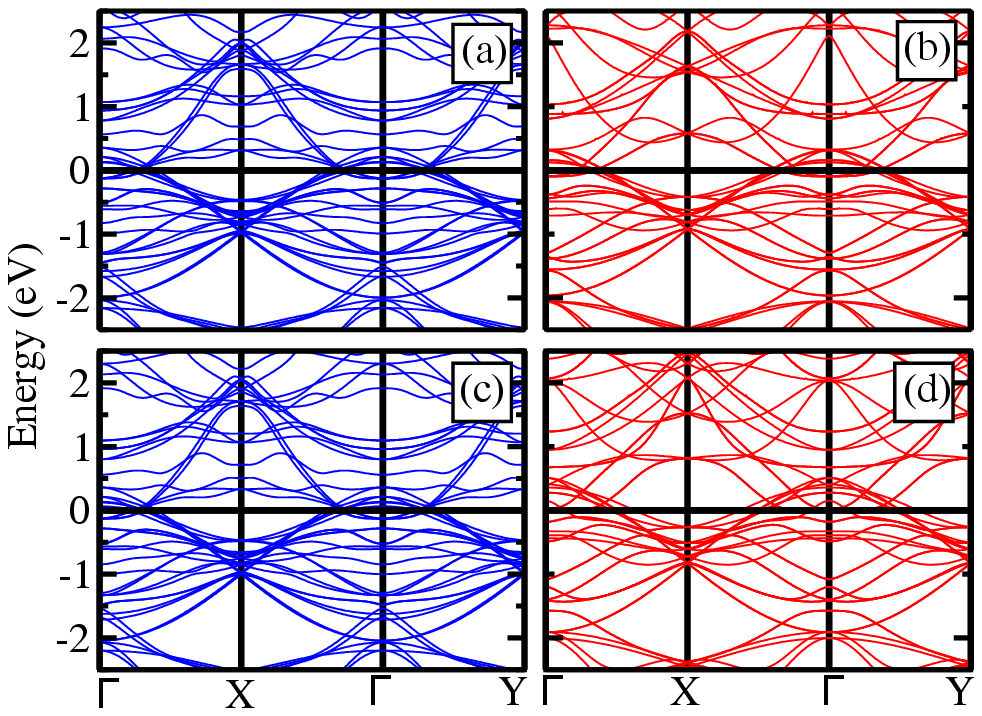}
  \caption{(Colour online) Calculated band structures (BS) of Ba$_{1-x}$K$_x$Fe$_2$As$_2$ (hole doping) 
  system near Fermi level (-2.5 eV to 2.5 eV) using SC (blue) methods for (a) $x=0.25$, (c) $x=0.5$ and VCA (red) methods
  for (b) $x=0.3$, (d) $x=0.5$.
  Horizontal solid black line at 0 eV indicates the Fermi level. Same k path ($\Gamma-X-\Gamma-Y$) has been
  chosen for BS calculation for two different methods.}
  \label{K_BS}
 \end{figure}
 Now we discuss the development of SC and VCA formalisms and some of their theoretical aspects. 
 The super-cell method defines a new unit cell with relatively large number of atoms where
 doping has been implemented, and then (conceptually) repeats 
 that cell throughout all space using periodic (or toroidal) boundary conditions \cite{Wannier}.
 For constructing our super-cells we follow three steps. At first we replicate the crystal unit cell 
 in all 3 directions to get a larger new unit cell with relatively large number of atoms.
 Then we introduce doping by substituting atoms with the dopant atoms. At the last step, we
 check the convergence of electronic properties with gradually increasing the size of the super-cell.
 On the other hand, the base of the VCA formalism is simple mixing of pseudopotential \cite{Ramer}.
 The supremacy of VCA method is its simplicity but in some cases it is not sufficiently accurate.
 The reason of incorrectness lies in the fact that only mixing of pseudopotential is considered.
 Ramer and Rappe \cite{Ramer} developed a more perfect VCA approach by considering 
 averages at the atomic level, such as the averages of eigenvalues of valence orbitals, charge
 densities of core electrons , nuclear potentials, wavefunctions {\it etc}. Major disadvantages of this method is
 that it is incapable of treating heterovalent atoms. In this work we use the VCA method, 
 developed by Bellaiche and Vanderbilt \cite{Bellaiche} based on weighted averaging of pseudopotentials.
 Below we describe the method briefly.
 Total energy of N valence electrons in terms of one-particle wave functions $\phi_i$ can be written in atomic units as:
 \begin{eqnarray}
 E_T[\{\phi_i\},\{{\bf R_l}\} ]=
 \sum\limits_{i}<\phi_i\mid-\frac{1}{2}\nabla^2+V_{ext}\mid\phi_i> \nonumber \\ 
 +\frac{1}{2}\int\int d{\bf r}d{\bf r'\frac{n({\bf r})n({\bf r'})}{\mid {\bf r}-{\bf r'}\mid}}+E_{XC}[n]+U(\{{\bf R_l}\})
 \end{eqnarray} 
 where
 \begin{eqnarray}
 V_{ext}({\bf r},{\bf r'})=\sum\limits_{l}V_{ps}^l({\bf r-R_l}, {\bf r'-R_l})             
 \end{eqnarray} 
 {\bf R$_{l}$} is the location of the site l, and $V_{ps}^l$ are the pseudopotentials. n({\bf r}) is the electron density
 at {\bf r}, $E_{XC}$ is the exchange correlation energy and $U(\{{\bf R_l}\})$ is the ion-ion interaction energy.
 In case of local as well as non local pseudopotential, 'VCA' operator equation can be derived by 
 averaging the  pseudopotential of the doped element on site l.
 \begin{eqnarray}
  V_{ps}^l({\bf r},{\bf r'})=(1-x)V_{ps}^A({\bf r},{\bf r'})+xV_{ps}^B({\bf r},{\bf r'})
  \end{eqnarray} 
  where, {\it e.g.,} A=Ba and B=K in Ba$_{1-x}$K$_x$Fe$_2$As$_2$. For other undoped sites C (Fe or As site), 
  pseudopotential can be written as,
  \begin{eqnarray}
    V_{ps}^l({\bf r},{\bf r'})=V_{ps}^C({\bf r},{\bf r'})
  \end{eqnarray} 
    Now V$_{ext}$ takes the form,
   \begin{eqnarray}
   V_{ext}({\bf r},{\bf r'})=\sum\limits_{l}\sum\limits_{S}w_S^lV_{ps}^S({\bf r-R_{lS}}, {\bf r'-R_{lS}})
   \end{eqnarray}
   where V$_{ps}^S$ and w$_S^l$ are the pseudopotential and 'weight' (statistical composition) of
   an atom of type {\it S} respectively.
 \begin{figure}[ht]
  \centering
  \includegraphics [height=4cm,width=8cm]{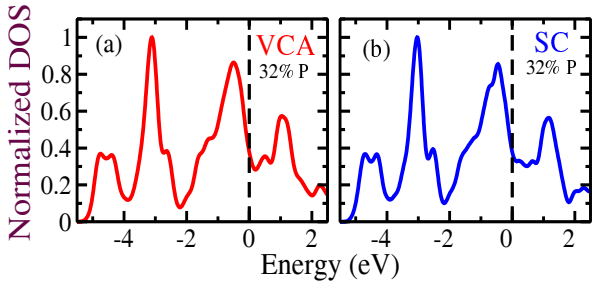}
  \caption{(Colour online) Calculated normalized total density of states using SC approach (blue solid line)
  as well as VCA approach (red solid line) for (a,b) BaFe$_2$(As$_{1-x}$P$_x$)$_2$ (isovalent doping) system with $x=0.32$
  within the energy interval -5.5 eV to 2.5 eV (near Fermi level, indicated by vertical dotted line at 0 eV).}
  \label{P_dos}
 \end{figure}
\begin{figure}[ht]
  \centering
  \includegraphics [height=7cm,width=8cm]{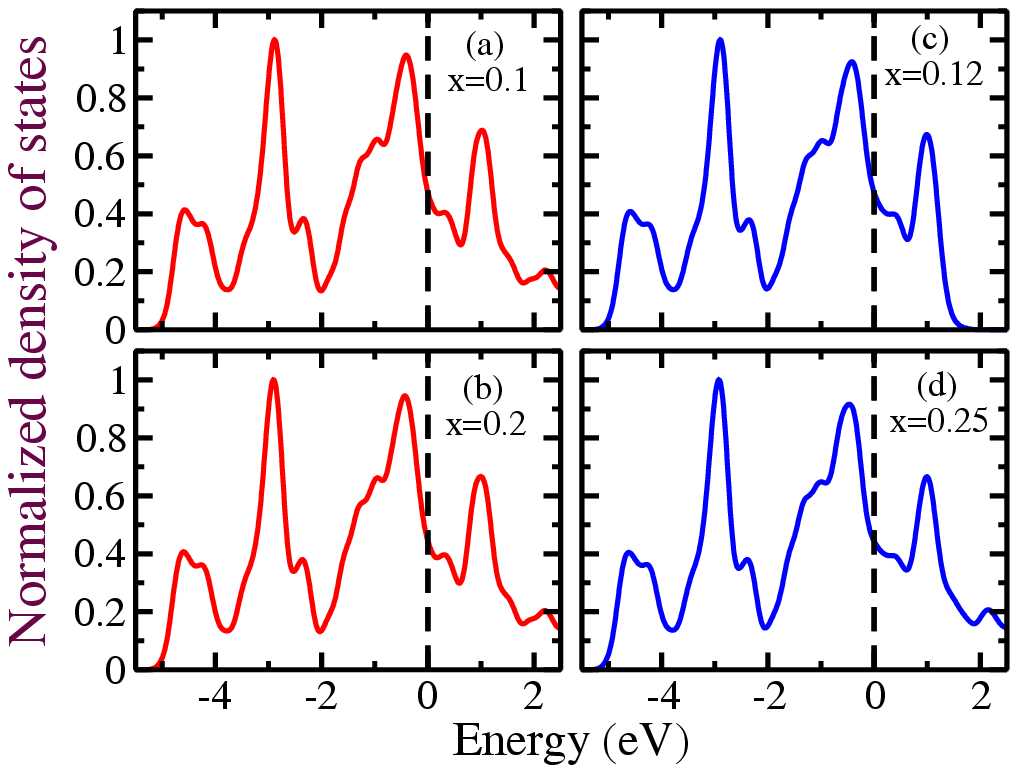}
  \caption{(Colour online) Calculated normalized total density of states for BaFe$_{2-x}$Co$_x$As$_2$ (electron doping) system
  using VCA approach (red solid line) with (a)$x=0.1$, (b)$x=0.2$ as well as SC approach (blue solid line) 
  with (a)$x=0.12$, (b)$x=0.25$
  within the energy interval -5.5 eV to 2.5 eV (near Fermi level, indicated by vertical dotted line at 0 eV).}
  \label{Co_dos}
 \end{figure}
\begin{figure}[ht]
  \centering
  \includegraphics [height=4cm,width=8cm]{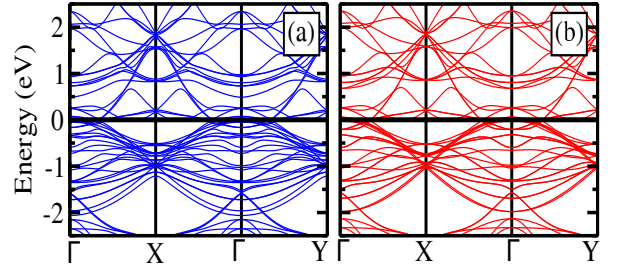}
  \caption{(Colour online) Calculated band structures (BS) of BaFe$_{2-x}$Co$_x$As$_2$ (electron doping) 
       system near Fermi level (-2.5 eV to 2.5 eV) using (a) SC (blue)($x=0.25$) and (b) VCA (red)($x=0.2$) methods.
       Horizontal solid black line at 0 eV indicates the Fermi level. Same k path ($\Gamma-X-\Gamma-Y$) has been
       chosen for BS calculation for two different methods.}
  \label{Co_BS}
 \end{figure}
\section{Result and Discussion}
We first calculate the orbital projected density of states (DOS) of BaFe$_2$As$_2$ system using 
experimental tetragonal lattice parameters $a$, $b$, $c$ and 
$z_{As}$ \cite{Rotter}. In FIG. \ref{PDOS_undoped} calculated 
partial density of states (PDOS) as well as total DOS near Fermi level (-6 eV to 6 eV)
 are presented. Contributions of various orbitals namely s, p and d to the total DOS 
 are depicted by blue, green and red solid lines respectively and dotted line indicates 
 the total DOS of the system. Fermi energy (E$_F$) is designated by the vertical dotted line at $0$ eV.
 It is very clear from FIG. \ref{PDOS_undoped} that $d$-orbital of Fe has the major contribution 
 to the total DOS near Fermi level. We also observed a significant contribution of $p$-orbital 
 near Fermi level, mainly coming from As $p$-orbital, to the total DOS. These observations 
 are very much consistent with earlier experimental and theoretical findings
 \cite{Graser,Zhang,Daghofer,Ran,Ding,Wang}. As mentioned earlier that pure BaFe$_2$As$_2$ system
 is not superconducting at ambient pressure but doping (at any of the sites) is one the possible 
 routes to make it a superconductor. In this work through detail first principles 
 electronic structures simulation, we provide a comparative study of electronic structures of 
 various doped BaFe$_2$As$_2$ systems, where implementation of doping has been done 
 by two different theoretical approaches: Virtual crystal approximation (VCA) and Super-cell (SC) methods.
\begin{figure}[ht]
  \centering
  \includegraphics [height=7cm,width=8cm]{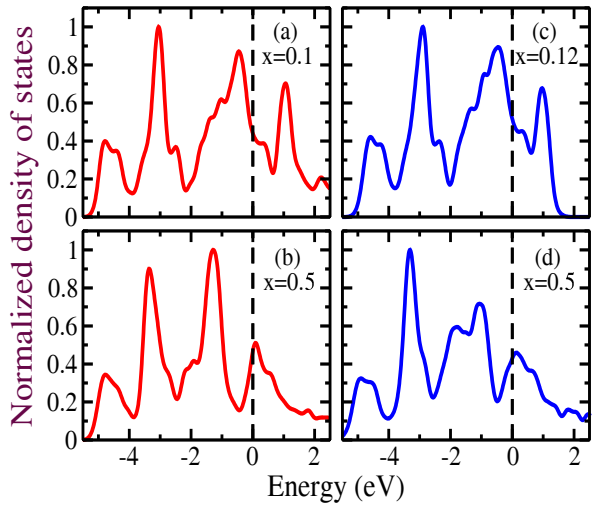}
  \caption{(Colour online) Calculated normalized total density of states for BaFe$_{2-x}$Ni$_x$As$_2$ (electron doping) system
      using VCA approach (red solid line) with (a)$x=0.1$, (b)$x=0.5$ as well as SC approach (blue solid line) 
      with (a)$x=0.12$, (b)$x=0.5$
      within the energy interval -5.5 eV to 2.5 eV (near Fermi level, indicated by vertical dotted line at 0 eV).}
  \label{Ni_dos}
 \end{figure}
 We take five doped BaFe$_2$As$_2$ systems 
 to examine the evolution of DOS and band structure (BS) with doping for both VCA and SC methods as mentioned previously.
 In FIG. \ref{K_dos_vca} and FIG. \ref{K_dos_sc} evolution of calculated total normalised DOS of 
 Ba$_{1-x}$K$_x$Fe$_2$As$_2$ system near Fermi level as a function of doping concentration has been 
 shown for VCA (red) and SC (blue) approaches respectively.
 \begin{figure}[ht]
   \centering
   \includegraphics [height=4cm,width=8cm]{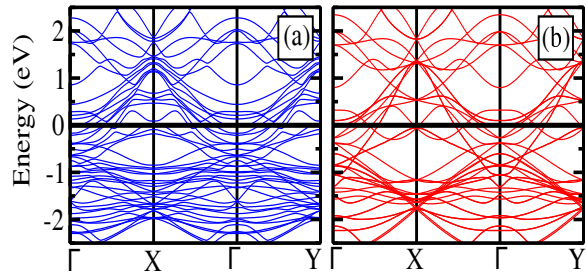}
   \caption{(Colour online) Calculated band structures (BS) of BaFe$_{2-x}$Ni$_x$As$_2$ (electron doping) 
     system with $x=0.5$ near Fermi level (-2.5 eV to 2.5 eV) using (a) SC (blue) and (b) VCA (red) methods.
     Horizontal solid black line at 0 eV indicates the Fermi level. Same k path ($\Gamma-X-\Gamma-Y$) has been
     chosen for BS calculation for two different methods.}
   \label{Ni_BS}
   \end{figure}
 \begin{figure}[ht]
  \centering
  \includegraphics [height=9cm,width=8cm]{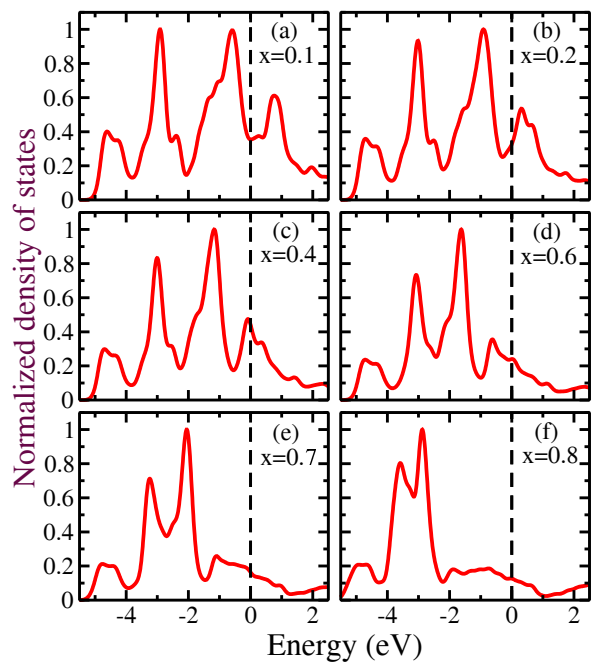}
  \caption{(Colour online) Calculated normalized total density of states (red solid line) using VCA approach
    for BaFe$_{2-x}$Ru$_x$As$_2$ (isovalent doping) system with (a) $x=0.1$, (b) $x=0.2$, (c) $x=0.4$,
    (d) $x=0.6$, (e) $x=0.7$ and (f) $x=0.8$
    within the energy interval -5.5 eV to 2.5 eV (near Fermi level, indicated by vertical dotted line at 0 eV).}
  \label{Ru_dos_vca}
 \end{figure}
 Calculated band structure of Ba$_{1-x}$K$_x$Fe$_2$As$_2$ systems 
near Fermi energy has been shown for $x=0.25$ and $x=0.5$ using SC approach 
in FIG. \ref{K_BS}a and FIG. \ref{K_BS}c respectively. Same has been presented using VCA formalism
with $x=0.3$ and $x=0.5$ in FIG. \ref{K_BS}b and FIG. \ref{K_BS}d respectively.
BS calculation has been performed along same k path for both the approaches (VCA and SC) 
to compare the electronic structures of these two cases. Calculated BS and DOS for Ba$_{1-x}$K$_x$Fe$_2$As$_2$ system
reveal that both the method (VCA and SC) produce very similar electronic structures. Since DOS 
around Fermi level of BaFe$_2$As$_2$ system is mainly Fe $d$-orbital derived, K doping in place of Ba does not 
affect the electronic structure near Fermi level significantly (with increasing K doping 
concentration FSs become more two dimensional \cite{pla}). As a result, K doping only shifts the chemical potential
 indicating the appearance of extra hole to the system which is reflected 
 in the calculated as well as experimentally measured FSs \cite{pla,Xu,Dhaka}. 
 But DOS profile as well as BS is almost unaffected 
 by K doping, in contrast to the cases of doping at other sites. 
 This is the main reason of getting same electronic structures for VCA and SC approaches.
 Central message from this study is that instead of SC method one can use time saving and numerically simpler VCA method for 
 studying the electronic structures such as BS, DOS, Fermi surface {\it etc.} of hole doped BaFe$_2$As$_2$ systems. 
 \begin{figure}[ht]
  \centering
  \includegraphics [height=9cm,width=8cm]{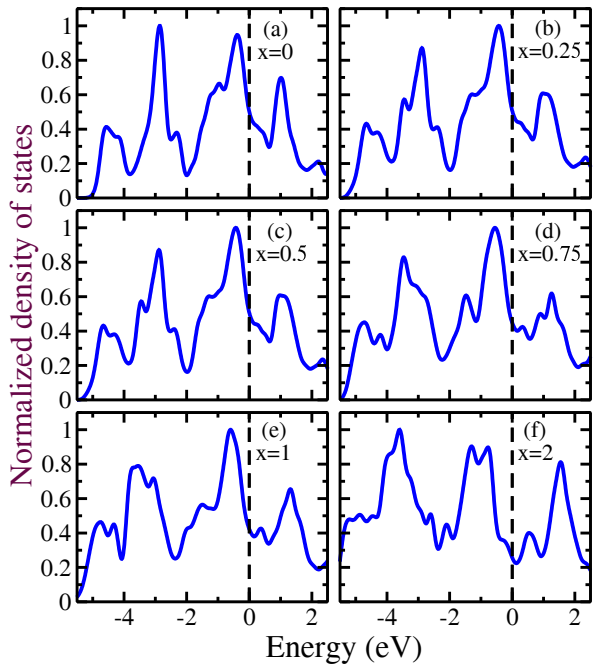}
  \caption{(Colour online) Calculated normalized total density of states (blue solid line) using SC approach
      for BaFe$_{2-x}$Ru$_x$As$_2$ (isovalent doping) system with (a) $x=0$, (b) $x=0.25$, (c) $x=0.5$,
      (d) $x=0.75$, (e) $x=1$ and (f) $x=2$
      within the energy interval -5.5 eV to 2.5 eV (near Fermi level, indicated by vertical dotted line at 0 eV).}
  \label{Ru_dos_sc}
  \end{figure}
 \begin{figure}[ht]
  \centering
  \includegraphics [height=7cm,width=8cm]{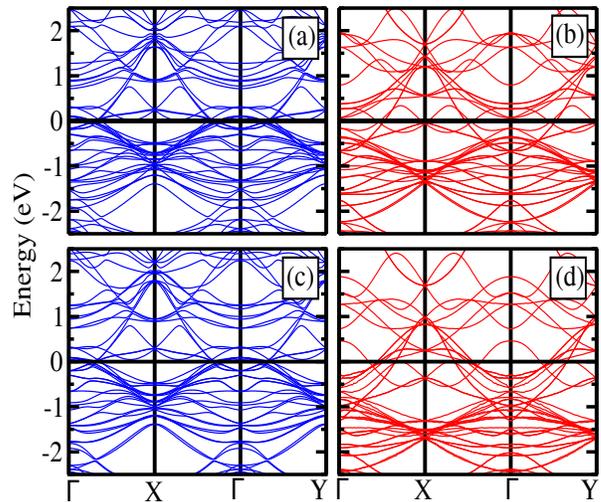}
  \caption{(Colour online) Calculated band structures (BS) of BaFe$_{2-x}$Ru$_x$As$_2$ (isovalent doping) 
   system near Fermi level (-2.5 eV to 2.5 eV) using (a,c) SC (blue) method with $x=0.25$ and $x=0.5$, 
   (b,d) VCA method (red) with $x=0.2$ and $x=0.5$.
   Horizontal solid black line at 0 eV indicates the Fermi level. Same k path ($\Gamma-X-\Gamma-Y$) has been
   chosen for BS calculation for two different methods.}
  \label{Ru_BS}
  \end{figure}
Next we have studied the case of isovalent P doping in place of As for BaFe$_2$As$_2$ system.
In FIG. \ref{P_dos}a and FIG. \ref{P_dos}b we present the normalized DOS of 32$\%$ 
P doped BaFe$_2$As$_2$ systems for SC (blue) as well as VCA (red) approaches respectively. Apart from 
nominal contrast in DOS profile very near to the Fermi level, we observed hardly any difference between the 
DOS calculated from SC and VCA methods. As with P doping in As site effectively modifies the Fe-As hybridization,
 one should expect some moderation of DOS around E$_F$, but unaltered large Fe-$d$ orbital contribution 
 to the total DOS restricts any notable modification of total DOS near Fermi level. As far as the two distinct methodologies 
 are concerned, we found both the approaches produce very similar DOS profile within the energy interval -5.5 eV 
 to 2.5 eV for 32$\%$ P doped BaFe$_2$As$_2$ systems. As discussed earlier DOS near E$_F$ is dominated by Fe 
 $d$-orbital, thus methods of implementation (VCA and SC) of P doping in As site play insignificant role in 
 the calculation of total DOS. Later we will discuss this issue to shed more light on the differences in the calculated electronic structures by these two approaches, providing the results of calculated PDOS of the 32$\%$ P 
 doped BaFe$_2$As$_2$ system.
 \begin{figure}[ht]
  \centering
  \includegraphics [height=4cm,width=8cm]{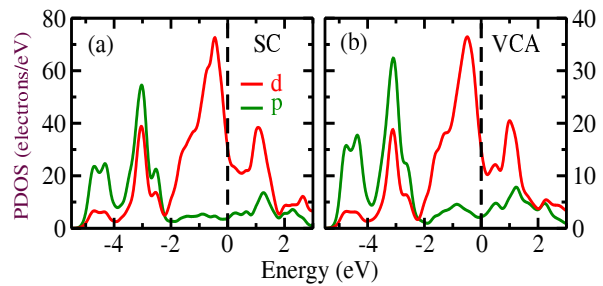}
  \caption{(Colour online) Calculated d (red) and p (green) projected partial density of states (PDOS)
  near Fermi level (-5.5 eV to 2.5 eV) for BaFe$_2$(As$_{1-x}$P$_x$)$_2$ (isovalent doping) system
   with $x=0.32$ using (a) SC and (b) VCA approaches.}
  \label{P_PDOS}
  \end{figure}
Now we study the case of doping on Fe site which is very
interesting as electronic structure near the Fermi level is dominated by Fe-$d$ orbitals.
First, we study the case of Co and Ni doping in place of Fe which exerts 
additional electron to the system and chemical potential shifts accordingly, 
is referred as electron doping. Phase diagram of electron doped (Co/Ni) Fe-based
superconductors clearly suggests the presence of superconducting as well 
as various other phases like SDW, Nematic {\it etc.} 
It should be noted that in case of electron doping, superconducting regime (also other phases) extended upto 15 $\%$
Co/Ni doping concentration. So study of electronic structure in the low doping regime
is well enough to draw conclusion about superconducting as well as other exotic normal state
 properties involving electronic structures. For electron doped system, we have 
 presented two case studies: (1) BaFe$_{2-x}$Co$_x$As$_2$ systems and (2) BaFe$_{2-x}$Ni$_x$As$_2$ systems.
 In FIG. \ref{Co_dos}a and FIG. \ref{Co_dos}b, we display the simulated normalized 
 DOS of BaFe$_{2-x}$Co$_x$As$_2$ systems with $x=0.1$ and $x=0.2$ and in FIG. \ref{Ni_dos}a and FIG. \ref{Ni_dos}b, 
 we depict the calculated normalized DOS of BaFe$_{2-x}$Ni$_x$As$_2$ 
 systems with $x=0.1$ and $x=0.5$, where doping has been implemented through VCA method. 
 FIG. \ref{Co_dos}c and FIG. \ref{Co_dos}d manifest the normalized DOS for 
 BaFe$_{2-x}$Co$_x$As$_2$ systems with $x=0.12$ and $x=0.25$, calculated using SC approach. 
 Same has been presented for BaFe$_{2-x}$Ni$_x$As$_2$ systems with $x=0.12$ and $x=0.5$ in 
 FIG. \ref{Ni_dos}c and FIG. \ref{Ni_dos}d. In FIG. \ref{Co_BS} 
 we demonstrate the calculated BS of BaFe$_{2-x}$Co$_x$As$_2$ systems with $x=0.2$ 
 for VCA (red) and with $x=0.25$ for SC (blue) methods. FIG. \ref{Ni_BS}a and FIG. \ref{Ni_BS}b represents the
 simulated BS of BaFe$_{2-x}$Ni$_x$As$_2$ systems with $x=0.5$ for SC and VCA approaches respectively.
It is evident from calculated BS as well as calculated DOS 
(FIG. \ref{Co_dos}c, FIG. \ref{Ni_dos}d, FIG. \ref{Co_BS}, FIG. \ref{Ni_BS}) that VCA and SC approaches produce
very similar electronic structure near E$_F$ in the low doping regime but
dissimilarities in the electronic structure arises in case of large doping like 50 $\%$ Ni doped BaFe$_2$As$_2$ system. 
In this case calculated DOS using VCA method has four distinct peaks 
within the energy interval -5.5 eV to 2.5 eV (see fig FIG. \ref{Ni_dos}b).
Where as calculated DOS implementing SC approach has five peaks (see fig FIG. \ref{Ni_dos}d) and DOS profile 
very near to the Fermi level is completely different from that of the calculated by VCA method.
In SC approach 50 $\%$ doping has been accomplished by substituting 50 $\%$ of the 
Fe atoms by Ni atoms from various random lattice sites of the bigger unit cell that we call super-cell. 
Consideration of two distinct sites for Ni and Fe in SC method is the main reason of the occurrence of 
two peak near Fermi energy instead of one. In VCA method 50 $\%$ doping has been implemented by mixing 
Fe and Ni pseudopotential in equal weightage. That is why one gets one peaks in the DOS 
 profile in VCA approach at a particular energy as a result of averaged peudopotential of Ni and Fe.
 Later we will show through PDOS calculation, in SC method Ni and Fe peaks appear at slightly 
 different energies on contrary to the case of calculated DOS by VCA method.
 Calculated BS (50 $\%$ Ni doped BaFe$_2$As$_2$ system) 
 for the two approaches appeared to be very similar as far as position of bands are concern 
 but a large no of extra band arises due to implementation of 
 Ni atom in distinct lattice sites in SC calculation (see FIG. \ref{Ni_BS}).
\begin{figure}[ht]
  \centering
  \includegraphics [height=7cm,width=8cm]{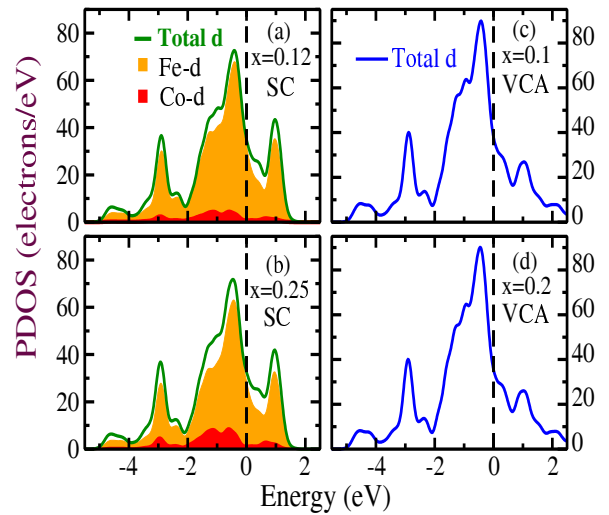}
  \caption{(Colour online) Calculated partial density of states (PDOS)
    near Fermi level (-5.5 eV to 2.5 eV) for BaFe$_{2-x}$Co$_x$As$_2$ (electron doping) system
   using SC method with (a) $x=0.12$ and (b) $x=0.25$ indicating contribution of Co-d (red) and Fe-d (orange) orbitals as well 
   as total $d$-orbital (green) to total DOS and VCA method with (c) $x=0.1$ and (d) $x=0.2$ indicating PDOS of $d$-orbital (blue).}
  \label{Co_PDOS}
  \end{figure}
 \begin{figure}[ht]
  \centering
  \includegraphics [height=7cm,width=8cm]{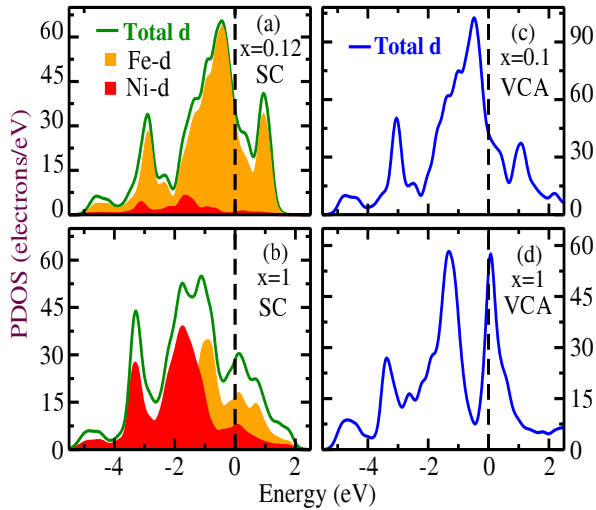}
  \caption{(Colour online) Calculated partial density of states (PDOS)
    near Fermi level (-5.5 eV to 2.5 eV) for BaFe$_{2-x}$Ni$_x$As$_2$ (electron doping) system
   using SC method with (a) $x=0.12$ and (b)$x=0.5$indicating contribution of Ni-d (red) and Fe-d (orange) orbitals as well 
   as total $d$-orbital (green) to total DOS and VCA method (c) $x=0.1$ and (d) $x=0.5$ indicating PDOS of $d$-orbital (blue).}
  \label{Ni_PDOS}
  \end{figure}

Next we study the case of isovalent doping on Fe site {\it i.e.,} Ru substitution in place of Fe.
FIG. \ref{Ru_dos_vca} demonstrates the calculated normalised DOS of Ru doped BaFe$_2$As$_2$ system 
as a function of Ru concentration where doping has been implemented via VCA formalism. On the other hand,
in FIG. \ref{Ru_dos_sc} we have presented the simulated normalized DOS for various Ru concentrations 
considering SC approach for the execution of Ru doping. FIG. \ref{Ru_BS} depicts the band structures of 
Ru doped BaFe$_2$As$_2$ system for two different Ru doping concentrations using both 
the two formalisms: VCA (red) and SC (blue). 
From FIG. \ref{Ru_BS} one can see that the position of the bands are not same for two different formalisms.
It is very clear from these figures (FIG. \ref{Ru_dos_vca}, FIG. \ref{Ru_dos_sc}, FIG. \ref{Ru_BS})
that the simulated DOS and BS are very much different for VCA and SC formalisms. It appears that in case of
VCA approach the chemical potential shifts a bit for larger doping concentration and deviates from the 
electronic structures calculated using SC method. A more detailed inspection of electronic structure
is thus required to unfold the reason of dissimilarities between the electronic structures calculated by VCA
and SC formalism. Below we present our results of calculated PDOS 
of all the BaFe$_2$As$_2$ systems that we discussed earlier
to get a better insight of the electronic structures calculated by these two approaches. 
\begin{figure}[ht]
  \centering
  \includegraphics [height=9cm,width=8cm]{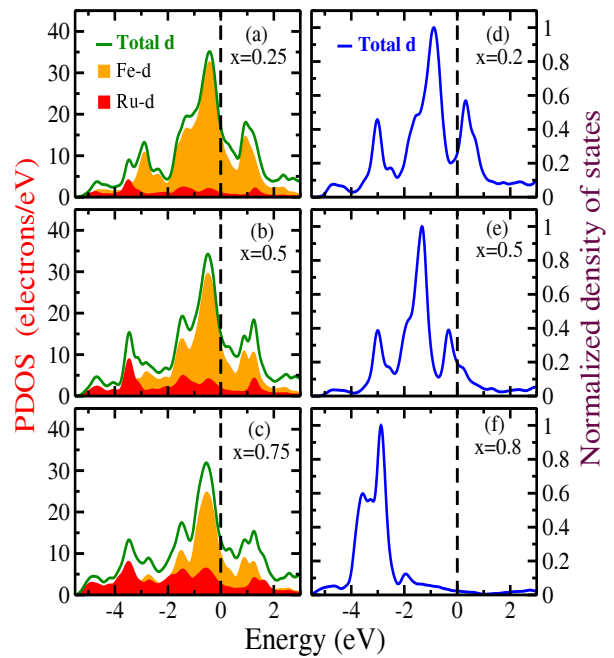}
  \caption{(Colour online) (a,b,c) Calculated partial density of states for BaFe$_{2-x}$Ru$_x$As$_2$ 
  (isovalent doping) system with $x=0.25$(a), $x=0.5$(b) and $x=0.75$(c) using SC method. Contribution of 
  Fe-d (orange), Ru-d (red) and total d (green) orbitals are indicated. (d,e,f) Normalized PDOS of $d$-orbital (blue)
  for BaFe$_{2-x}$Ru$_x$As$_2$ (isovalent doping) system with $x=0.2$(d), $x=0.5$(e) and $x=0.8$(f) using VCA method.}
  \label{Ru_PDOS}
 \end{figure}
\begin{figure}[ht]
  \centering
  \includegraphics [height=14cm,width=8cm]{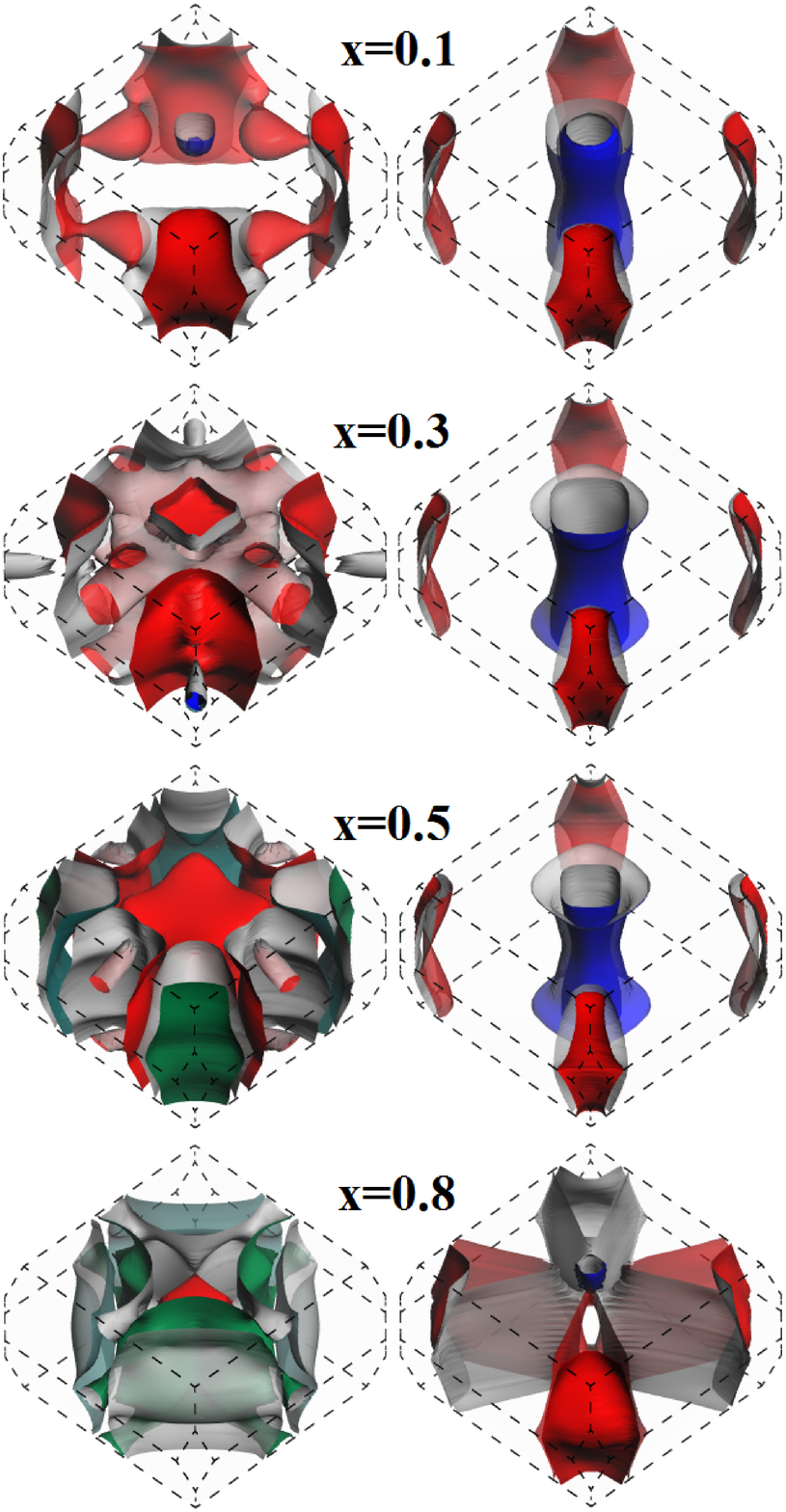}
  \caption{(Colour online) Simulated Fermi surfaces of BaFe$_{2-x}$Ru$_x$As$_2$ using VCA methods with out any chemical shift (left column) and with proper chemical shift (right column). Chemical shift is estimated by comparing calculated DOS for VCA and SC approaches.}
  \label{FS}
 \end{figure}

Since K doping at Ba site does not alter the DOS much, near the Fermi level, it is irrelevant to
show the PDOS of these K doped systems or more precisely hole doped systems in the context of our 
main aim of this work. First we see the differences in calculated PDOS for VCA as well as SC methods in case of 32$\%$
P doped BaFe$_2$As$_2$ system. FIG. \ref{P_PDOS} depicts the calculated orbital projected PDOS for
the two approaches. Total $d$ and $p$ orbital contributions are presented for both the cases.
P doping in As site effectively modifies the PDOS of $p$-orbital which in turn results some amount of moderation of
Fe-As hybridization. From FIG. \ref{P_PDOS}, it is evident that both the approaches simulate same 
PDOS of $d$-orbital but there is remarkable differences in PDOS of $p$-orbital calculated by two different methods.
The differences are more pronounced if one compares the PDOS of $p$-orbital with respect to the PDOS of $d$-orbital 
for the two formalism, that we considered. Now we proceed to the case of Co/Ni doping in Fe site.
In FIG. \ref{Co_PDOS} and FIG. \ref{Ni_PDOS} we depict the
calculated PDOS of BaFe$_{2-x}$Co$_x$As$_2$ and BaFe$_{2-x}$Ni$_x$As$_2$ systems respectively.
In the left column of both the Figures, results of SC calculation is presented, where PDOS of 
Fe-d (orange), Ni/Co-d (red) as well as total $d$-orbitals (solid green line) are shown for various doping
concentrations as indicated in the figures. Calculated total $d$-orbital contribution (solid blue line) using
VCA method is also displayed in the right column of both the figures (see FIG. \ref{Co_PDOS} and FIG. \ref{Ni_PDOS})
for comparison with that of the calculated PDOS using SC methods (left column).
Comparative study of calculated PDOS of Co/Ni doped systems for both the methods reveals that in case of 
50$\%$ Ni doping VCA and SC method produce significantly different electronic structures.
One can see from FIG. \ref{Ni_PDOS}b that the peak of the 
$d$-orbitals of Ni and Fe are at slightly different energies and sum of these two DOS profiles produce 
the total $d$-orbital contribution. But in case of VCA, PDOS of the total $d$-orbital as shown in FIG. \ref{Ni_PDOS}d 
is totally different from that 
of the SC calculation.
Although implementation of higher doping is not that much important in case of electron doped
 system as discussed earlier but one must concern about the fact that in case of higher doping concentration 
 VCA method is unable to simulate the accurate electronic structures.
Next we show in FIG. \ref{Ru_PDOS}, the simulated PDOS of Ru doped BaFe$_2$As$_2$ system for VCA and SC methods.
 In SC calculation (left) we presented the $d$-orbital contribution of Ru-d (red) as well as Fe-d (orange)
 separately and total $d$ orbital contribution displayed by solid green line (see 
 FIG. \ref{Ru_PDOS}a, b, c). On the right side of FIG. \ref{Ru_PDOS} total $d$-orbital contribution 
 has been depicted for VCA method. Comparison of both the electronic structure reveals that
 in VCA formalism, mixed $d$-orbital of Ru and Fe atom deviates from the PDOS of total $d$-orbital calculated by SC method.
 Although Ru doping at the Fe site is iso-electronic substitution but total number of electron as well as size of Ru 
 is much greater than that of Fe. Furthermore Ru has $4d$ orbital as the outermost orbital, whereas outermost orbital of 
 Fe is 3d. It also appears that in the low doping concentration, apart from an additional shift 
 of the chemical potential in the calculated electronic structures by VCA method, there is not 
 much difference in the electronic structures calculated by these two formalisms. But in case of higher doping 
 concentrations, electronic structure simulated using VCA formalism deviates remarkably from that of the SC method. 
 Perhaps in VCA formalism, the mixing of 3d (Fe) orbitals with 4d (Ru) orbitals which are more broader than 3d,
 is not implemented properly resulting an additional shift in the chemical potential 
 in DOS profile for VCA approach.
 This discrepancy can be overcome in the low doping regime.
 In FIG. \ref{FS} (left column) we depict the calculated FSs of Ru doped BaFe$_2$As$_2$ systems
 using VCA formalism. In the right column, we present the FSs of the same by giving
 an additional shift , which is estimated from the comparison of calculated DOS by VCA and
 SC methods. It is clear from FIG. \ref{FS} that in the low doping regime the shifting of chemical potential
 work very nicely to produce the FSs, that resembles with the experimentally measured 
 FSs as well as calculated FSs using SC methods \cite{pla,Xu,Dhaka}. 
 But in case of higher Ru doping concentration this shifting of chemical potential 
 does not work to reproduce the FSs accurately as evident from the simulated FS 
 in the last row of FIG. \ref{FS}. Calculated FSs using VCA formalism in case of 40$\%$ 
 Ru doping (last row of FIG. \ref{FS}) certainly not resembles with that of the SC calculation 
 on 50$\%$ Ru doped systems \cite{pla}.\\

\section{Conclusion}
Implementing VCA and SC methods for doping, we have presented a detail comparative study 
of electronic structures of various doped BaFe$_2$As$_2$ systems through first principles 
simulation. We have investigated the density of states (including partial density of states), 
band structures and Fermi surfaces of electron doped, hole doped as well as isovalently 
doped BaFe$_2$As$_2$ systems for VCA and SC approaches. In case of hole doping and iso-electronic 
P substitution on As site, both the methods produce very similar electronic structures as far as 
DOS and BS calculations are concerned. Calculated PDOS of this system 
give clear indication about the fact that P doping in As site (isovalent) moderate Fe-As hybridization.
So in this case there is some difference in the calculated PDOS for VCA and SC approaches.
In case of electron doped (Co/Ni) BaFe$_2$As$_2$ systems VCA and SC methods provide very 
similar electronic structures for both the cases for low doping concentrations. 
But in case of higher doping concentration, we find that the calculated electronic structures are 
remarkably different for these two methods as clearly evident from the BS and PDOS calculation of 
50$\%$ Ni doped BaFe$_2$As$_2$ systems. On the other hand, in case of iso-electronic substitution 
in Fe site (Ru substitution in Fe site), calculated electronic structure is very much dissimilar for 
different methods. An extra shift in the chemical potential is observed in the electronic structures of 
Ru doped BaFe$_2$As$_2$ systems, when VCA method is employed for doping.
But for higher Ru doping concentration the electronic structures are quite different for these two methods. 
We also explicitly show that by considering this chemical shift in a proper way, one can simulate 
FSs accurately by VCA formalism in case of low Ru doping concentration. 
This work give a clear idea about the application of VCA and SC methods for calculating the electronic 
structure of various doped 122 systems of Fe-based superconductors.

\section{Acknowledgements} We thank Dr. P. A. Naik and Dr. P. D. Gupta for their encouragement in this work. One of us (SS) acknowledges the HBNI, RRCAT for financial support and encouragements. 

\end{document}